\documentclass[preprint2]{aastex}

\shorttitle{Collisionless model of the solar wind}
\shortauthors{Zouganelis et al.}

\begin{document}

\title{A transonic collisionless model of the solar wind}

\author{I. Zouganelis\altaffilmark{1}, M. Maksimovic\altaffilmark{1}, N. Meyer-Vernet\altaffilmark{1}, H. Lamy\altaffilmark{2} and K. Issautier\altaffilmark{1}}

\altaffiltext{1}{LESIA, Observatoire de Paris, 5 place Jules
Janssen, 92195 Meudon, France;\\ ioannis.zouganelis@obspm.fr}
\altaffiltext{2}{Belgian Institute for Space Aeronomy, 3 Avenue
Circulaire, B-1180 Brussels, Belgium}

\begin{abstract}
Because of the semi-collisional nature of the solar wind, the
collisionless or exospheric approach as well as the hydrodynamic
one are both inaccurate. However, the advantage of simplicity
makes them useful for enlightening some basic mechanisms of solar
wind acceleration. Previous exospheric models have been able to
reproduce winds that were already nearly supersonic at the
exobase, the altitude above which there are no collisions. In
order to allow transonic solutions, a lower exobase has to be
considered, in which case the protons are experiencing a
non-monotonic potential energy profile. This is done in the
present work. In this model, the electron velocity distribution in
the corona is assumed non-thermal. Parametric results are
presented and show that the high acceleration obtained does not
depend on the details of the non-thermal distributions. This
acceleration seems, therefore, to be a robust result produced by
the presence of a sufficient number of suprathermal electrons. A
method for improving the exospheric description is also given,
which consists in mapping particle orbits in terms of their
invariants of motion.
\end{abstract}


\section{Introduction} \label{introduction}

Most cosmic bodies eject matter into space, but the solar wind is
the first and, up to now, only stellar outflow to have been
measured in situ \citep{neu97}. Numerous sophisticated models have
been developed since Parker's pionnering papers \citep{par58,
par60} using complicated mechanisms. The different acceleration
mechanisms, the origin of the high-speed solar wind as well as the
associated problem of coronal heating have been recently reviewed
in a comprehensive way by \citet{cra02}. However the solar wind
acceleration and its properties are still far from being well
understood. The reason of this difficulty is that the solar wind
is neither a collision-dominated medium nor a collisionless one.
The Knudsen number $K_n$, which is defined as the ratio of the
particle mean free path and the density scale height, is close to
unity at Earth's orbit (see e.g. \citet{hun72}). This means that
neither the hydrodynamic approach nor the pure collisionless one
(also called exospheric) are fully appropriate to model the solar
wind expansion and to explain its observed properties.

The classical fluid approach is applicable for the extreme regime
when $K_n\ll1$, that is when the medium is collision-dominated. In
this case, the particle velocity distribution functions (VDFs) are
Maxwellians as the medium is assumed to be at local thermodynamic
equilibrium. The Euler or Navier-Stokes approximations are
applicable and produce a thermally driven wind out of the hot
solar corona. There are two problems with this approach. Firstly,
the particle VDFs might not be Maxwellians at the base of the
solar wind. Secondly, the particle VDFs $are$ $not$ Maxwellian in
the solar wind. There is an increasing number of both theoretical
\citep{vin00, leu02}, and observational evidences \citep{ess00},
which tend to show that non-thermal VDFs can develop and exist in
the high corona and even in the transition region. This is because
in a plasma the particle free paths increase rapidly with speed
$(\propto v^4)$, so that high energy tails can develop for Knudsen
numbers as low as $10^{-3}$ \citep{sho83}, i.e. even in
collisional plasmas. A fortiori, high energy tails can be expected
to be found in the weakly collisional corona and solar wind
acceleration region. Indeed, it is well known that the solar wind
electron VDFs permanently exhibit non-thermal tails that can be
modelled by a halo Maxwellian population (for e.g. \citet{fel75})
or by the power law part of a generalized Lorentzian or Kappa
function \citep{mak97b}. In the frame of the fluid approach, which
intrinsically cannot handle suprathermal tails, the effect of
non-thermal VDFs on the solar wind acceleration can be understood
through an increase of the heat flux \citep{hol78, olb81}.

An alternative way of taking into account the possible effects of
coronal non-thermal distributions is to use a kinetic approach.
Among the various kinetic approaches for the solar wind, the
simplest one is probably the exospheric one, which totally
neglects binary collisions between particles above a given
altitude called the exobase. The first solar wind model of this
type was developed by \citet{cha60} by analogy with the
evaporation of planetary atmospheres. This first exospheric model,
modelling the radial expansion of the solar corona from the
thermal evaporation of hot coronal protons out of the solar
gravitational field, produced a solar breeze. The subsonic speed
obtained by the theory was partially due to an inadequate
assumption: the electrostatic field was taken so as to ensure
hydrostatic equilibrium \citep{pan22, ros24}, which is
inconsistent with an expanding atmosphere. The improved exospheric
models by \citet{jen63} and \citet{bra66} were the first to be
able to reproduce supersonic solar wind flows. In these models,
multiple exobase locations were assumed, in order to take into
account the variable mean free path of the particles as a function
of their velocity. However these models still used the inadequate
Pannekoek-Rosseland electric field as an imposed external
solution. The actual (outward) ambipolar electric field, which
ensures plasma quasi-neutrality and zero electric current is
greater, thereby accelerating protons to greater speeds. Models
using this correction \citep{lem71a, joc70} produced supersonic
winds, but too small speeds for explaining the fast solar wind
$(\sim 700-800\ km\ s^{-1})$.

More recently, \citet{mak97a} have generalized these calculations
 by considering non-Maxwellian velocity distribution functions for
the electrons in the corona, e.g. generalized Lorentzian or Kappa
functions. With such non-Maxwellian distributions having
suprathermal electron tails, a higher electrostatic potential is
needed to ensure zero charge and current, therefore producing
larger terminal bulk speeds. In essence, this comes about because
the electron tail tends to increase the escaping electron flux, so
that, to preserve quasi-neutrality, the electrostatic potential
increases in order to trap more electrons, which in turn
accelerates the protons outwards. This model yields a reasonable
description of bulk solar wind properties, giving densities,
temperatures and speeds within the ranges observed at 1 AU, even
though the details of the VDFs are not reproduced, as expected
since collisions are neglected. Its major interest is the
prediction of high speeds without assuming extremely large coronal
temperatures and/or additional heating of the outer corona, as is
needed in hydrodynamic models. More basically, the main
achievement of exospheric models is to furnish a possible driving
mechanism for the fast solar wind, with a single assumption: the
suprathermal electron VDF at the exobase.

However, the \citet{mak97a} model cannot be applied for low
altitude exobases, which is the case of coronal holes, from where
emanates the fast solar wind. In these deep coronal layers, the
gravitational force acting on the protons is stronger than the
electric one, so that the total potential for the protons is
attractive out to some distance where the two forces balance each
other. Farther out, the outward electric force dominates. This
means that the total potential energy for the protons is not
monotonic, presenting a maximum at a certain distance from the
exobase \citep{joc70} and therefore not all the protons present at
the exobase are able to escape. The presence of such a maximum is
not taken into account in the \citet{mak97a} model, nor in the
\citet{lem71a} one, since both models started beyond the vicinity
of the sonic point.

The purpose of the present paper is to study the effect of
non-thermal electron VDFs in the frame of a transonic exospheric
solar wind model. We use a special technique described by
\citet{joc70} and \citet{kha98} consisting in mapping particle
orbits in terms of the invariants of motion. The same problem has
been recently considered by \citet{lam03} using a different
formulation, which involves an approximation relative to the
escaping particles rate (see Appendix \ref{emucalcul}). The
present work, which uses the same approximation, sets the basis of
an exact exospheric description of the solar wind acceleration.
The reader should, however, have in mind that the present model is
still a very simplified one, which does not pretend to describe
all solar wind properties. Instead, it may be useful to determine
some basic aspects and a possible driving mechanism of the solar
wind, avoiding ad hoc assumptions on energy dissipation and using
as few free parameters as possible.

In section \ref{basics} we recall the basics of an exospheric
solar wind model. In section \ref{protons} we outline the
difficulties arising when dealing with a non-monotonic potential
energy for the protons and describe the technique used to
calculate the interplanetary electrostatic potential. In section
\ref{electrons} we deal with non-Maxwellian electron distribution
functions and consider three cases: a Kappa, a sum of two
Maxwellians (a core and a halo) and a sum of a Maxwellian and a
Kappa. The latter distribution is rather general and reproduces
the main features of VDFs observed in space plasmas: a Maxwellian
profile at low speeds and a high energy tail with a power law
shape. In section \ref{results} we describe the results of our
model and compare them with observations. A summary and final
remarks are given in section \ref{conclusion}.

\section{Basics of a kinetic collisionless model} \label{basics}

In exospheric or kinetic collisionless models, a specific
altitude, called the exobase, is defined as an abrupt boundary
between the collision dominated region (in which a hydrodynamic
approach is valid) and a completely collisionless one. This
boundary is usually defined as the distance $r_0$ from the center
of the Sun where the Knudsen number $K_n$ is equal to unity, i.e.
where the Coulomb mean free path becomes equal to the local
density scale height. Although the mean free path depends on the
particle temperature and, therefore the exobase should be
different for protons and electrons, a common exobase is assumed
in order to simplify the calculations. Furthermore, in order to
concentrate on the basic physics and since the exobase location is
defined from parameters that are not accurately known, we put the
exobase at the surface of the Sun, i.e. $r_0=1R_{\odot}$. This
stands as an approximation for a more realistic value that should
be slightly greater and has no important qualitative impact on the
results.

The collisionless nature of the plasma above the exobase allows
one to calculate the velocity distribution of each particle
species as a function of the distribution at the exobase by using
Liouville's theorem with conservation of energy $E$ and magnetic
moment $\mu$ in order to solve the Vlasov equation for a time
stationary wind. These two constants of motion are defined as:
\begin{equation}\label{energy}
    E=\frac{mv^2}{2}+m\phi_g+Ze\phi_E
\end{equation}
\begin{equation}\label{magnmom}
    \mu=\frac{mv_{\perp}^2}{2B}
\end{equation}
where $v$ is the speed of the particle of mass $m$, $Ze$ its
charge, $\phi_g(r)=-M_\odot G/r$ the gravitational potential,
$\phi_E(r)$ the interplanetary electrostatic potential and
$v_{\perp}$ the velocity component perpendicular to the magnetic
field vector $\vec{B}$. The choice of using the constants of
motion $(E,\mu)$ instead of the speed and pitch angle is
convenient as it removes the spatial dependence in the
distribution function. In a purely collisionless model, the
magnetic moment is conserved, even though this may not be true for
protons in the high-speed wind \citep{sch83}.

The classification of particles in different species is based on
their trajectories along magnetic field lines. These trajectories
depend on the velocity and pitch angle at the exobase, or
correspondingly on their energy and magnetic moment. There are
four different classes of trajectories: incoming, escaping,
ballistic and trapped particles, as defined in \citet{lem71a}.
Escaping are the particles that have enough energy to escape from
the Sun, while ballistic are those with insufficient energy that
are returning towards it. Another kind of non-escaping particles
are the trapped ones, which do not have enough energy to escape,
but whose inclination to the magnetic field lines is large enough
that they are reflected by the magnetic mirror force before
reaching the exobase $r_0$. Finally there are particles coming
from infinity, called incoming, which are neglected. Indeed, due
to the postulated absence of collisions above the exobase, no
particles, in principle, are backscattered in the downward loss
cone. This assumption can be relaxed in future applications of our
model, but it is not unreasonable in the fully collisionless case.

Once a VDF $f_0$ is assumed at the exobase level $r_0$, the VDF
$f$ at any larger distance $r$ is uniquely determined by
Liouville's theorem. The electron and proton distributions $f_e$
and $f_p$ and their moments - in particular the electron and
proton densities $n_e$ and $n_p$ - are functions of the electric
potential $\phi_E(r)$. The quasi-neutrality condition
$n_e(r)=n_p(r)$ is then used to determine, by an iterative method,
the value of the potential $\phi_E$ at any altitude $r$. The
iterative process is stopped when the estimation of $\phi_E(r)$ is
adequate within the required precision. In addition we impose the
equality of the electron and proton fluxes at the exobase, in
order to ensure a zero electric current. (Note that the current
will also be zero everywhere because fluxes of both kinds of
particles vary as $r^{-2}$). Then, it is possible to calculate the
other moments, e.g. pressures and heat flux, at any distance $r$.
Strictly speaking, one should solve Poisson's equation instead of
imposing quasi-neutrality. However, for scales much greater than
the Debye length, which is the case here, the quasi-neutrality
condition provides a very good approximation for the electrostatic
potential distribution in the solar wind.

Another assumption is the use of a radial interplanetary magnetic
field $B(r)$, varying as $r^{-2}$, which means that the rotation
of the Sun as well as a possible super-radial expansion of the
wind are neglected. As was shown by \citet{che72} with a fluid
approach and more recently by \citet{pie01} with an exospheric
approach using Kappa distributions, a spiral magnetic field does
not change significantly the wind density and bulk speed. However,
temperatures are modified and especially their anisotropies. In
any case the magnetic field is not very far from radial up to
about $1AU$ so that this approximation is reasonable in the
framework of the present model in which we study fundamental
aspects of the fast solar wind coming from coronal holes, i.e.
rather high heliolatitudes. Nevertheless, a super-radial expansion
of the wind is an ingredient that might influence the results, but
it is neglected in this zero order approximation in order to avoid
additional free parameters.

\section{Non-monotonic proton potential energy} \label{protons}

Solar wind protons are subjected to the attractive gravitational
potential $\phi_g(r)$ of the Sun and to the repulsive
interplanetary electrostatic field $\phi_E(r)$. Their total
potential energy is thus $\Psi(r)=m\phi_g(r)+e\phi_E(r)$, which
reaches a maximum at some radial distance $r_{max}$. Such a
maximum occurs because the electric force is expected to decrease
slower than $r^{-2}$, so that this force, which pushes the protons
outwards, should dominate gravity at large distances \citep{joc70,
mey03}. Below this altitude $r_{max}$, the gravitational force is
larger than the electrostatic one and the total potential is
attractive. The opposite is true above $r_{max}$, forcing all the
protons present at these altitudes to escape.

Until now, exospheric models using suprathermal distributions for
the electrons have considered a monotonic potential profile for
the protons. This means that the exobase was located above the
maximum, i.e. at $r_0>r_{max}$, so that the wind was already
supersonic at the exobase level. All protons were escaping, as
they were experiencing a monotonic repulsive total potential. Even
though these models gave supersonic solutions, they were
incomplete because they did not consider the transition from
subsonic to supersonic speeds. Note that in the general case the
Parker's critical point does not coincide with the potential
maximum $r_{max}$ and is generally below it \citep{mey03}, even
though these two points are relatively close to each other. This
is also true in recent kinetic simulations taking into account
Coulombian collisions \citep{lan03}. In the special case of an
increasing temperature with $T\propto r^{3/4}$, these two points
coincide \citep{scu96}.

For coronal holes, from where emanates the fast solar wind, the
relation $r_0>r_{max}$ is no longer valid. Indeed, in that case,
the exobase is located deeper in the corona, because of the lower
density of coronal holes. At these altitudes, some protons cannot
escape from the gravitational well of the Sun and become ballistic
or trapped. Such a case has been considered by \citet{joc70}, who,
however did not consider non-thermal distributions for the
electrons. Recently, \citet{lie98} have given a more general
theoretical framework for dealing with arbitrary potential
profiles, while \citet{kha98} have applied it to describe the
precipitation of a magnetospherically trapped hot population and
the outflow from the high-latitude ionosphere.

With a non-monotonic potential, the validity of previous
exospheric models is questionable because of the violation of
constraints given by \citet{chi78} regarding the first and second
derivatives of the electrostatic potential with respect to the
magnetic field. The technique described by \citet{lie98} consists
in mapping particle orbits in terms of the invariants of motion.
This technique has the interest of removing the spatial dependence
of the distributions, which can now be written as $f(E,\mu)$. The
density, flux and pressures are then given by the following
expressions:
\begin{equation}\label{densemu}
    n=\frac{\sqrt{2}\pi B}{m^{3/2}}\int\frac{f(E,\mu)}{\sqrt{E-\mu
    B-\Psi}}dEd\mu
\end{equation}
\begin{equation}\label{fluxemu}
    F=\frac{2\pi B}{m^2}\int f(E,\mu)dEd\mu
\end{equation}
\begin{equation}\label{pressparalemu}
    P_{\parallel}=\left(\frac{2}{m}\right)^{3/2}\pi B\int\sqrt{E-\mu B-\Psi}f(E,\mu)dEd\mu
\end{equation}
\begin{equation}\label{pressperpemu}
    P_{\perp}=\frac{\sqrt{2}\pi B^2}{m^{3/2}}\int\frac{\mu f(E,\mu)}{\sqrt{E-\mu
    B-\Psi}}dEd\mu
\end{equation}
The spatial dependence is now transferred into the region of
integration (over which the moments are calculated) in $E-\mu$
space. The basic problem is the accessibility of the different
particle populations in this space. This technique has been
recently outlined by \citet{zou03}; for a full detailed
description of the method, we refer the reader to the paper by
\citet{lie98}. In Appendix \ref{emucalcul}, we describe some
elementary concepts of this technique, as applied to the solar
wind.

\section{Non-thermal electron distributions} \label{electrons}

Maxwellian distributions can be used for the solar wind protons,
but solar wind electrons permanently exhibit significant
suprathermal tails in their distribution functions. The
fundamental reason for the electrons to be non-Maxwellian is that
fast electrons collide much less frequently than slow ones because
of their greater free path, so that they cannot relax to a
Maxwellian. Usually the solar wind electron distributions are
fitted to a sum of two Maxwellians: a core of cold electrons and a
halo of hot ones (see for instance \citet{fel75}, \citet{pil87}).
An alternative to the core/halo model has been proposed by
\citet{vas68}, that is a generalized Lorentzian or a Kappa
function, which has been introduced by \citet{scu92} to explain
high coronal temperatures by velocity filtration. An interesting
combination of these two distributions is the sum of a cold
Maxwellian and a hot Kappa as a halo component. The sum of two
Maxwellians is then a particular case of this last one.

\subsection{Kappa function} \label{kappasect}

The electron distributions in the solar wind are observed to have
important high velocity tails and a convenient way to fit them is
to use Kappa (or generalized Lorentzian) functions. This was done
by \citet{mak97b} using Ulysses data. Some recent observations
\citep{lin97} suggest the existence of a superhalo-Maxwellian
population that does not seem to be fitted by a Kappa function.
However, this part of the distribution $(v\gg v_{th})$ does not
contribute in our model. The main interest of Kappa distributions
is that they require one fewer parameter than the core/halo model,
whereas having a power-law suprathermal tail as often observed in
space \citep{vas68}. Recently, there have been some attempts to
find a physical explanation of these distributions \citep{col93,
ma99, tre99, leu02} and to study how non-thermal electron tails
can be generated in the lower chromosphere \citep{vin00} and the
corona \citep{voc03}.

\citet{mak97a} have considered for the first time Kappa
distributions in a kinetic collisionless model of the solar wind,
but with the restriction of a monotonic potential energy for the
protons, thus resulting in a wind that is already supersonic at
the exobase. The Kappa function is defined as:
\begin{equation}\label{kappadef}
    f_{\kappa}(\vec{v})=\frac{n_{e0}}{(\pi\kappa
    v_{\kappa}^2)^{3/2}}\frac{\Gamma(\kappa+1)}{\Gamma(\kappa-1/2)}\left(1+\frac{v^2}{\kappa
    v_{\kappa}^2}\right)^{-(\kappa+1)}
\end{equation}
where $\Gamma(x)$ is the Gamma function and $v_{\kappa}$ is the
thermal speed defined by:
\begin{equation}\label{vthermkappa}
    v_{\kappa}=\left(\frac{2\kappa-3}{\kappa}\frac{k_bT_e}{m_e}\right)^{1/2}
\end{equation}
where $k_b$ is the Boltzmann constant and $m_e$ the electron mass.

For speeds $v$ smaller or comparable to $v_{\kappa}$, the Kappa
distribution, for any value $\kappa\geq 2.5$, is rather close to a
Maxwellian having the same thermal speed. However, the equivalent
Kappa temperature $T_{\kappa}$ (defined from the second moment of
the VDF, as the ratio between pressure and density) is related to
the Maxwellian one $T_M$, by
$T_{\kappa}=(\kappa/(\kappa-3/2))T_M$. For $v\gg v_{\kappa}$, the
Kappa distribution decreases with $v$ as a power law
$(f_{\kappa}\varpropto v^{-2(\kappa+1)})$. In the limit
$\kappa\rightarrow\infty$, $f_{\kappa}(v)$ reduces to a Maxwellian
distribution with $T_{\kappa\rightarrow\infty}=T_M$. Note that
when electron distributions measured in the solar wind are fitted
with Kappa functions, the parameter $\kappa_e$ for the electrons
ranges from 2 to 5 \citep{mak97b}.

The basic equations for an exospheric model with Kappa
distributions are given by \citet{pie96}. Those concerning the
electrons are still valid in our model since the potential energy
of electrons is an increasing function of the distance. When
considering a Kappa VDF, we therefore use the equations (5)-(34)
by \citet{pie96}, which give the density, flux, pressure and
energy flux of the electrons above the exobase; these quantities
correspond respectively to the zero, first, second and third order
moments of the Kappa VDF.

\subsection{A sum of two Maxwellians} \label{maxwsumsect}

Kappa VDF is the most convenient function modelling distributions
with high energy tails. However, there are several reasons for
considering other kinds of non-thermal VDFs. Firstly, there is
currently no agreed-upon physical explanation for Kappa VDFs and,
even though they are rather close to a Maxwellian at low
velocities, they may not be as close to it as observed. Secondly,
moments of order higher than $2\kappa-1$ diverge since
$f_\kappa\propto v^{-2\kappa-2}$ at large speeds. Note however
that this is merely a mathematical difficulty, which has no impact
on our model since it does not involve these higher order moments.
Thirdly, from Liouville's theorem, the value of $\kappa$, which
represents the non-thermal character, remains constant with
altitude in the absence of collisions. Actually, one should expect
that as distance increases and there are less collisions, the
electron VDFs should present stronger suprathermal tails, so that
$\kappa$ should decrease with distance. Indeed, recent
observations in the corona suggest distributions having a
non-thermal character that increases with altitude \citep{ess00}.

For these reasons we also consider a sum of two Maxwellians, which
is the classical way of representing electron VDFs in the solar
wind \citep{fel75,pil87}. Such a VDF does not have any of the
above-mentioned disadvantages. If $n_{c0}$ and $T_{c0}$ are
respectively the electron density and temperature at the exobase
for the cold (core) component and $n_{h0}$ and $T_{h0}$ are the
same quantities for the hot (halo) component, we can define their
relative importance by two parameters: $\alpha_0=n_{h0}/n_{c0}$
and $\tau_0=T_{h0}/T_{c0}$. The total density and temperature at
the exobase is then given by: $n_0=n_{c0}(1+\alpha_0)$ and
$T_{e0}=((1+\alpha_0\tau_0)/(1+\alpha_0))T_{c0}$. At any altitude
the density is just the sum of the two densities, the core and the
halo one. The same is true for the fluxes and pressures. What is
interesting is that now the parameters $\alpha$ and $\tau$ are
functions of the distance, so that the non-thermal nature of the
distribution is not held constant contrary to the case of a Kappa
distribution. Note that in this purely collisionless model the
core and halo components do not interact with one another (as in
e.g. the hydrodynamic core-halo model of \citet{che03}). The
density, flux and pressures are defined as the moments of the
distribution function for each particle population (escaping,
ballistic and trapped) as given in \citet{pie96}. Analytical
expressions of these quantities can be calculated in function of
the electrostatic potential $\phi_E(r)$. For a Maxwellian VDF,
they are similar to those given in \citet{lem71b} and will not be
expressed here. These expressions are to be used for both the core
and the halo components.

\subsection{A sum of a Maxwellian and a Kappa function} \label{maxwkappasumsect}

A more general form for the distribution function is the sum of a
Maxwellian core and a Kappa halo. This function is closer to a
Maxwellian at low speeds than a Kappa distribution and has the
advantage that the non-thermal character is not held constant as
was the case with a Kappa VDF. We use the same definitions of
$\alpha_0$ and $\tau_0$ as before. Note that for
$\kappa\rightarrow\infty$ the VDF reduces to a sum of two
Maxwellians. The expressions of the moments used are those for a
Maxwellian for the core and the equations (5)-(34) by
\citet{pie96} for the Kappa halo.

\section{Results and Discussion} \label{results}

As explained in section \ref{basics}, the exobase location is
approximated to be at $r_0=1R_\odot$ with no serious impact on the
results. We assume for the temperatures at the exobase
$T_{e0}=10^6\ K$ and $T_{p0}=2T_{e0}$, in the range of values
observed in coronal holes \citep{cra02}. The density at the
exobase does not affect the results of velocity or temperature and
is just a multiplicative factor in the density profiles.

\begin{figure} \plotone{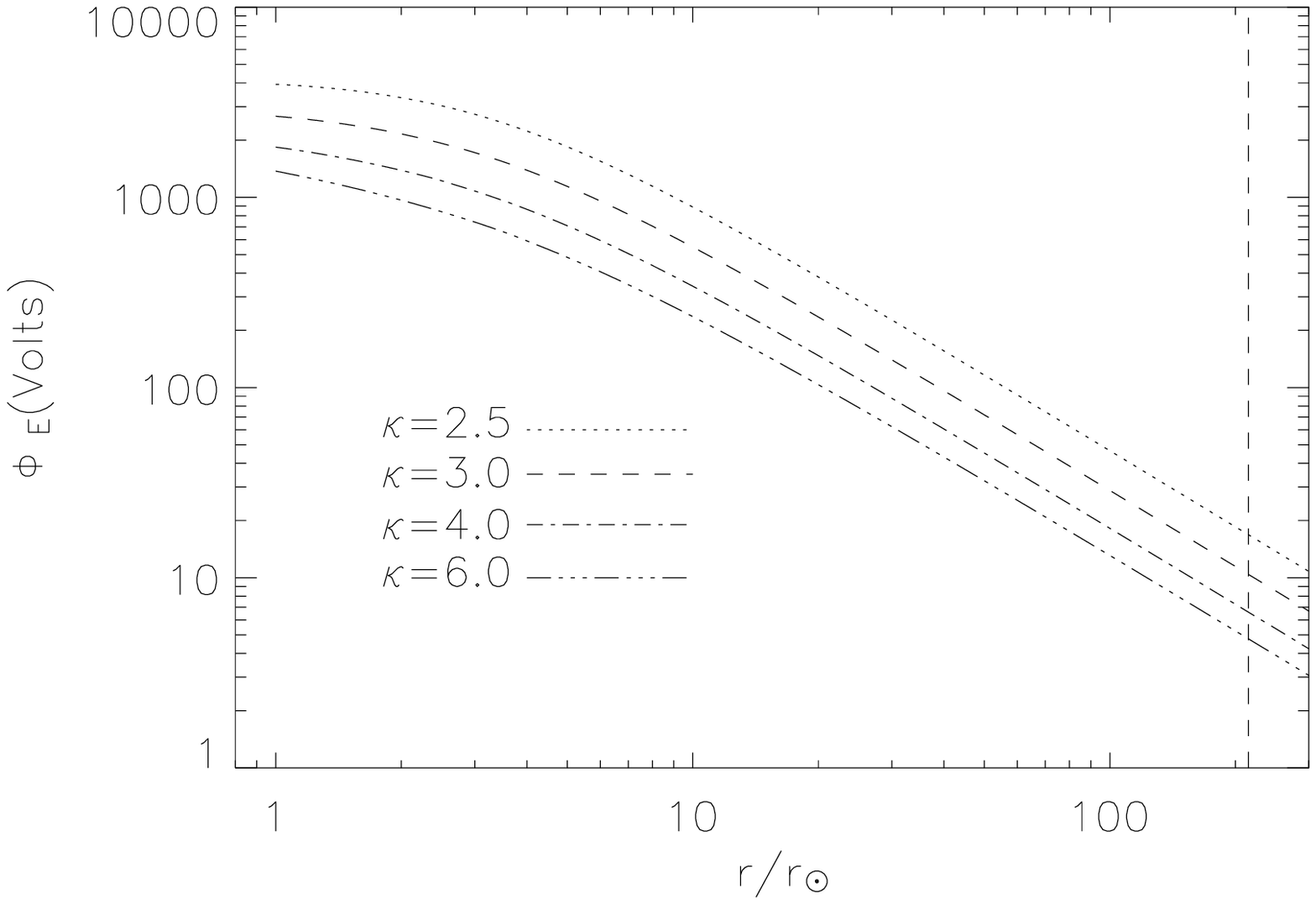} \caption{Interplanetary electrostatic potential for different values of $\kappa=2.5$, $3.0$, $4.0$ and $6.0$. The dashed vertical line indicates Earth's orbit. \label{phikappa}}
\end{figure}

\begin{figure} \plotone{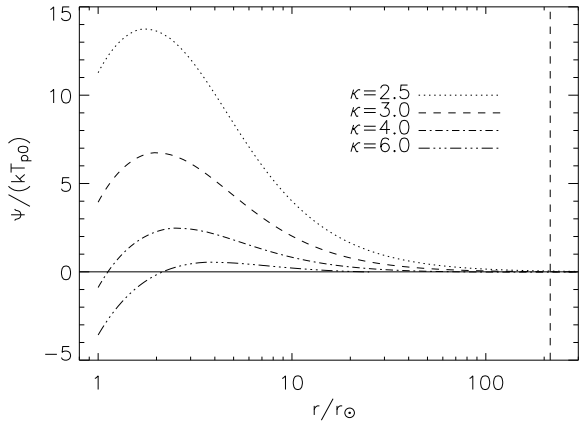} \caption{Total proton potential energy for different values of $\kappa=2.5$, $3.0$, $4.0$ and $6.0$. The dashed vertical line indicates Earth's orbit. The energy is normalized to $k_bT_{p0}$. \label{psikappa}}
\end{figure}

Let us first consider a Kappa VDF for the electrons as described
in section \ref{kappasect}. The calculated electric potential and
the total potential energy of the protons are plotted in Figures
\ref{phikappa} and \ref{psikappa} respectively for different
values of $\kappa$ ranging from $\kappa=6$ to $\kappa=2.5$, a case
with a conspicuous suprathermal tail. Note that we use $\kappa>2$
in order for the energy flux to be finite. One sees that the value
of the maximum of potential increases and its distance $r_{max}$
decreases as $\kappa$ decreases. This is because with more
suprathermal electrons, a stronger electric potential is needed to
preserve quasi-neutrality. For a Maxwellian VDF
$(\kappa\rightarrow\infty)$, the total proton potential energy
increases monotonically and tends asymptotically to zero
(remaining always negative).

\begin{figure} \plotone{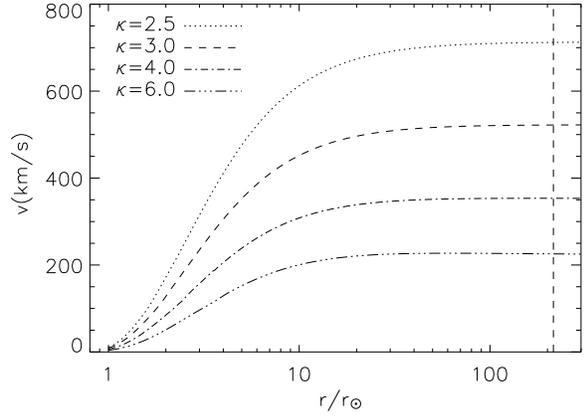} \caption{Bulk speed profiles for $\kappa=2.5$, $3.0$, $4.0$ and $6.0$. The dashed vertical line indicates the Earth's orbit. \label{vitkappa}}
\end{figure}

The bulk speed - the ratio between flux and density - is shown in
Figure \ref{vitkappa} for a Kappa distribution. A high terminal
bulk speed $(>700\ km\ s^{-1})$ is obtained when the suprathermal
tail is conspicuous $(\kappa=2.5)$. This is due to the large value
of the maximum in ion potential energy $(\approx14k_bT_{p0})$,
which is transformed into kinetic energy of the escaping protons
as they are accelerated above $r_{max}$. An important remark is
that the major part of this high terminal bulk speed is obtained
within a small heliocentric distance $(\approx10R_\odot)$; this is
due to the large acceleration represented by the large slope of
the potential above $r_{max}$. Note that this is the largest
terminal bulk speed obtained by this model with a Kappa VDF.

\begin{figure} \plotone{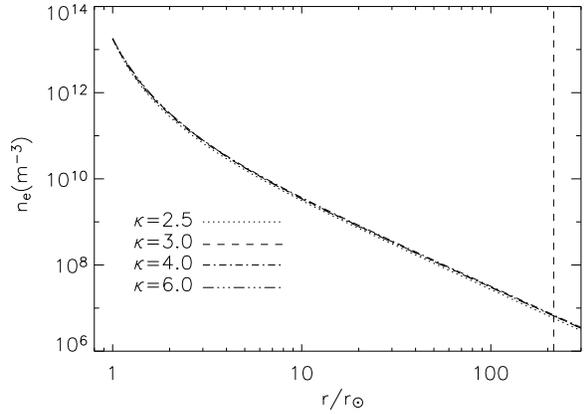} \caption{Electron density profiles for $\kappa=2.5$, $3.0$, $4.0$ and $6.0$. The dashed vertical line indicates the Earth's orbit. \label{densekappa}}
\end{figure}

The density profiles are shown in Figure $\ref{densekappa}$. One
sees that they are nearly independent of the value of $\kappa$.
The density at $1AU$ depends on the one taken at the exobase. In
this figure we assumed an exobase density $n_0=1.8\cdot 10^{13}\
m^{-3}$, which corresponds to the density of a coronal hole
extrapolated to $r_0=1R_{\odot}$ as given by \citet{kou77} in line
with recent studies on atmospheric and coronal electron densities
\citep{ess99}. With this density, the model yields a density of
about $6\ cm^{-3}$ at $1AU$, of the same order as observed in
situ. Note that the rest of the results do not depend on the
density.

This analysis bears out previous results of an exospheric model
using Kappa VDFs \citep{mak97a}. There are however two basic
differences. In the present work the velocity profiles span the
whole domain from the subsonic to the supersonic regime, which was
not the case when the exobase was located above $r_{max}$. The
second difference is that we obtain high bulk speeds with more
reasonable temperatures at the exobase. Note however that a direct
comparison cannot be done because of a slightly different exobase
definition resulting in different proton temperatures at the
exobase. In any case both models can produce high bulk speeds
without assuming an additional (ad hoc) heating mechanism in the
outer corona, as is generally postulated in hydrodynamic models.
Furthermore, the fact that a faster wind is obtained with low
values of $\kappa$ agrees with observations showing that VDFs have
large suprathermal tails in the fast solar wind but are closer to
a Maxwellian in the slow wind \citep{mak97b}.

\begin{figure} \plotone{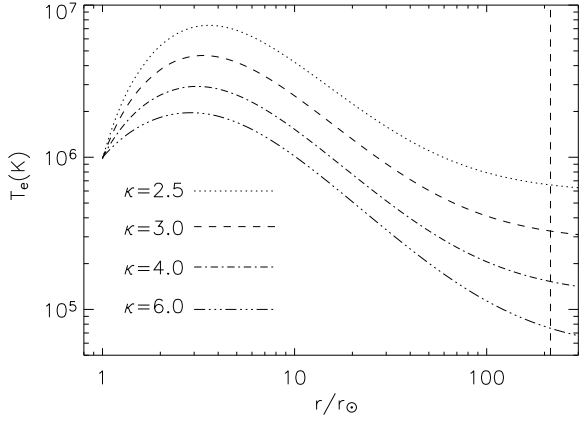} \caption{Electron temperature profiles for $\kappa=2.5$, $3.0$, $4.0$ and $6.0$. The dashed vertical line indicates the Earth's orbit. \label{tempekappa}}
\end{figure}

The large suprathermal tails for low values of $\kappa$ have
another important consequence. They make the electron temperatures
increase considerably with distance up to a maximum
$(\approx7$x$10^6\ K)$ within a few solar radii. This maximum in
electron temperature is smaller for larger values of $\kappa$ and
disappears as $\kappa\rightarrow\infty$ $(Fig.\ref{tempekappa})$
as does the maximum in the total potential energy of protons. This
temperature increase is a direct consequence of filtration of the
non-Maxwellian VDF by the attracting electrostatic potential
\citep{scu92}. This large temperature increase is not observed,
which suggests that Kappa functions may not be adequate to model
VDF having suprathermal tails in the corona.

\begin{figure} \plotone{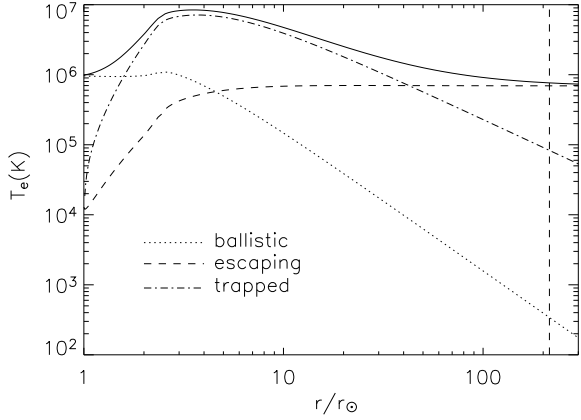} \caption{Electron temperature profile (full line) for a sum of a Maxwellian core and a Kappa halo with $\alpha=0.03$, $\tau=5$ and $\kappa=2.5$. The other lines show the contributions of the different particle species. The dashed vertical line indicates the Earth's orbit. \label{tempecontrib}}
\end{figure}

Let us now consider the results obtained with electron
distributions made of a sum of two Maxwellians or a sum of a
Maxwellian core and a Kappa halo. On the whole the results are
rather similar. For the same acceleration we obtain approximately
the same temperature increase as in the Kappa case as we can see
in Figure \ref{tempecontrib} $(\alpha_0=0.03, \tau_0=5$ and
$\kappa=2.5)$, which corresponds to a terminal bulk speed of
$\sim770\ km\ s^{-1}$. We deduce that the temperature increase is
not an artefact of Kappa VDFs, but a general behavior of
non-thermal distributions. For a given terminal bulk speed the
filtration mechanism results in the same temperature increase. It
is important to note that collisionless models are expected to
give correct electron temperatures \citep{mey98}, because
collisions with other particles do not significantly affect the
electron energy, whereas collisions between electrons do not
change their total temperature. In any case we should remind that
the present model is still a zero-order one, which is intended to
explore the basic physics of the wind acceleration, but should not
be expected to reproduce all observations in a detailed way,
because it involves very few free parameters.

Figure \ref{tempecontrib} shows also the contributions of the
different particle species to the total electron temperature. At
large distances the temperature profile is the sum of a term
$\propto r^{-4/3}$ plus a constant. The $r^{-4/3}$ term comes from
the isotropically distributed electrons (ballistic and trapped)
confined by the heliospheric electric potential, which is found to
have the same radial variation at large distances. The constant
term comes from the parallel temperature of the escaping
electrons. This agrees with analytical results by \citet{mey98}
that do not depend on the VDFs in the corona, but were obtained
with a monotonic proton potential profile. When the proton
potential energy is non-monotonic, the asymptotic electron
temperature profile is still the sum of a term varying as
$r^{-4/3}$ plus a constant \citep{mey03}, but the relative
importance of these terms is not necessarily the same.

\begin{figure} \plotone{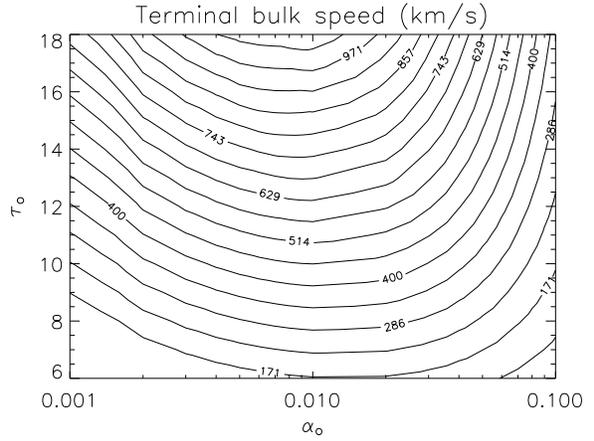} \caption{Contours of the terminal bulk speed (at 1AU) for a sum of two Maxwellians for the electrons as a function of $\alpha_0$ and $\tau_0$ at the exobase.\label{contourmm}}
\end{figure}

In Figure \ref{contourmm} we show results for a sum of two
Maxwellians. The diagram shows contours of the terminal bulk speed
(at 1AU) as function of $\alpha_0$ and $\tau_0$, where we can see
that this kind of VDF is able to explain the values of the fast
wind speed $(~700-800\ km\ s^{-1})$. The terminal bulk speed
increases with increasing $\tau_0$, which is not surprising since
the halo temperature increases. Concerning the parameter
$\alpha_0$, the terminal bulk speed behaves differently. One sees
that the terminal speed has a maximum for some value of $\alpha_0$
(for a given $\tau_0$). This is reasonable since for
$\alpha_0\rightarrow 0$ and $\alpha_0\rightarrow\infty$ we have
just one Maxwellian with temperature $T_{e0}$ and for all values
of $\alpha_0$ between these limits, the electron VDF is
non-thermal giving rise to a more important acceleration because
of the velocity filtration mechanism. In addition, close to these
limits the terminal bulk speed becomes independent of $\tau_0$ as
there is just only one VDF. That makes the contour lines to be
vertical.

\begin{figure} \plotone{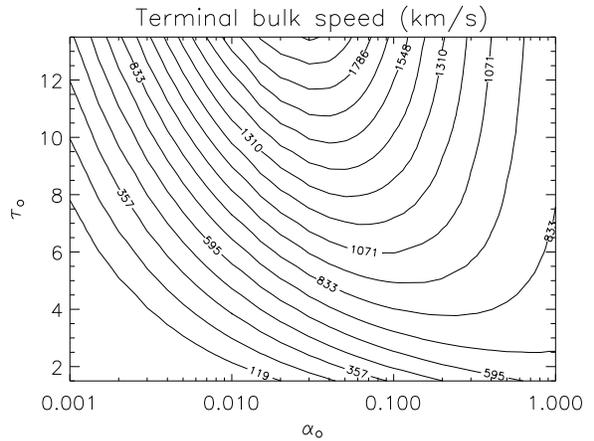} \caption{As in fig.$\ref{contourmm}$, but for a sum of a Maxwellian core and a Kappa halo with $\kappa=2.5$.\label{contourmk}}
\end{figure}

When using a sum of a Maxwellian core and a Kappa halo (for
instance with $\kappa=2.5$) the contour plot of the terminal bulk
speed is quite similar to the previous one as is shown in
Fig.$\ref{contourmk}$. The main difference is the higher
acceleration, which is due to the use of the Kappa function. The
maxima for a given $\tau_0$ are now displaced to the right (to
larger values of $\alpha_0$). This is due to the fact that for
$\alpha_0\rightarrow\infty$ the VDF is now just a single Kappa
accelerating the wind more than a single Maxwellian. It is
important to note that we can obtain very high wind speeds even
without using a Kappa VDF (Fig.\ref{contourmm}). This shows that
the acceleration is not just a consequence of the Kappa function,
but results from non-thermal distributions, as expected. There are
no restrictions on $\alpha_0$ and $\tau_0$ (as in the case of the
Kappa VDF where we have to take $\kappa>2$), but one should
constrain these parameters by coronal observations or by future in
situ measurements close to the Sun.

\section{Summary and Final Remarks} \label{conclusion}

In the present work we have described a collisionless model of the
solar wind acceleration assuming non-thermal velocity distribution
functions in the corona. The base altitude of the fast wind was
taken to be low enough (case of coronal holes), so as to consider
the transition from the subsonic to the supersonic regime. That
needed a special resolution method considering a non-monotonic
potential energy of the protons. An approximation regarding the
escaping particles rate has been used (see Appendix
\ref{emucalcul}) in order to obtain the electrostatic potential in
a self consistent way.

There are two important results in our work. Firstly, we have
shown the fundamental role of non-thermal electron velocity
distributions in accelerating the wind. The high value of the
terminal bulk speed is not just an artefact of the use of Kappa
functions. Such speeds can be obtained also with a sum of two
Maxwellians (a cold and a hot one), which is the most commonly
used model to represent the observed electron distributions.

Secondly, there is a more important acceleration of the wind
compared to previous exospheric models. This is due to the
non-monotonic proton potential profile that forces some of the
protons to return back to the Sun and, therefore, reduces the
escaping proton flux. As a consequence, the interplanetary
electrostatic potential accelerating the wind is enhanced. It is
also important to note that the terminal bulk speed is
anticorrelated with the ratio of proton to electron temperatures
at the base of the wind and therefore there is no need to assume
large coronal temperatures or additional heating of the outer
region of the corona in order to explain fast wind speeds.

An inherent property of non-thermal distributions is the velocity
filtration mechanism \citep{scu92}. This results in an increase of
the electron temperature within a few solar radii. One problem of
the present model is the relatively high maximum reached by the
electron temperature when using highly non-thermal electron
distributions at the exobase to produce the fast wind. This
problem could be due to the approximation used in order to obtain
the potential in a self consistent way. A full treatment of the
problem with no approximation would increase the terminal bulk
speed, but the influence on the electron temperature is not known
a priori. Another ingredient that might influence the results is a
possible super-radial expansion of the wind.

An apparent inconsistency of our model is the population of
trapped electrons. Trapped electrons can only be produced if
collisions decelerate ballistic electrons in order to be reflected
before reaching the exobase. Strictly speaking, in the present
fully collisionless model, trapped electron orbits should not be
populated. If so, however, the VDF would have a strong
discontinuity which should be rapidly smoothed out by even a very
small level of collisions. Hence, we think that it is more
reasonable - as was done in all previous exospheric models - to
set this population at a level ensuring that the VDF has no
discontinuity for non-escaping particles.

The collisionless nature of our model is an inherent drawback.
Recent kinetic simulations of the solar wind taking into account
binary collisions between particles \citep{lan03} suggest that
collisions might be an important ingredient for accelerating the
wind to supersonic speeds, but this latter work does not consider
non-thermal electron distributions. The role of collisions has
also been studied by \citet{lan01} and \citet{dor03} for the solar
corona. The electron heat flux, which plays a key role in the
solar wind acceleration, seems to be essentially determined by the
collisionless high-energy tail. This suggests that even though the
present model neglects collisions, it may correctly describe a
large part of the physics involved in the effect of non-thermal
electron distributions. In order to better explain the
observational properties, additional physical ingredients should
be taken into account, which will be the purpose of a future
study.

\acknowledgments

We are grateful to A. Mangeney, M. Moncuquet and F. Pantellini for
stimulating discussions and useful comments on the manuscript. We
also thank the anonymous referee for an amount of helpful
suggestions that improved this paper.

\appendix

\section{Protons accessibility in the $E-\mu$ phase space} \label{emucalcul}

\begin{figure} \plotone{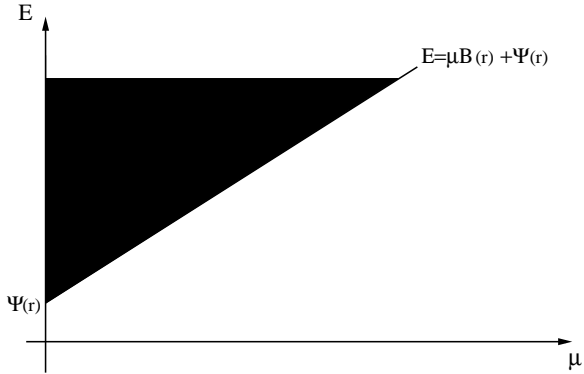} \caption{$E-\mu$ space for
the altitude $r$. The slope of the $r$ limit line is the amplitude
of the magnetic field $B(r)$, while its intercept is the total
proton potential energy $\Psi(r)$ at this altitude. The
distribution function $f$ is defined only above this limit line,
i.e. in the black region. \label{emusimple}}
\end{figure}

In this appendix we recall some elementary concepts of the
resolution method when dealing with protons in a non-monotonic
potential energy structure. The conservation laws (\ref{energy})
and (\ref{magnmom}) determine the region where the function $f$ is
defined as:
\begin{equation}\label{khazinequalitydefinitionregion}
    v_{\parallel}^2\geq0\Rightarrow E\geq\mu B(r)+\Psi(r)
\end{equation}
The relation (\ref{khazinequalitydefinitionregion}) defines the
line $v_{\parallel}=0$ for each altitude $r$; the distribution
function $f$ is defined only above this line, as is shown in
Figure $\ref{emusimple}$. From now on we will call this line at
the altitude $r$, as the '$r$ limit line'. Note that the slope of
this line is just the amplitude of the magnetic field $B(r)$
(noted from now on as $B_r$). Since in our case, the magnetic
field is always decreasing, the sharpest limit line will be the
one of the reference level $r_0$. Note also that the intercept of
an $r$ limit line corresponds to the total proton potential energy
$\Psi(r)$ at this altitude (also noted as $\Psi_r$).

\begin{figure} \plotone{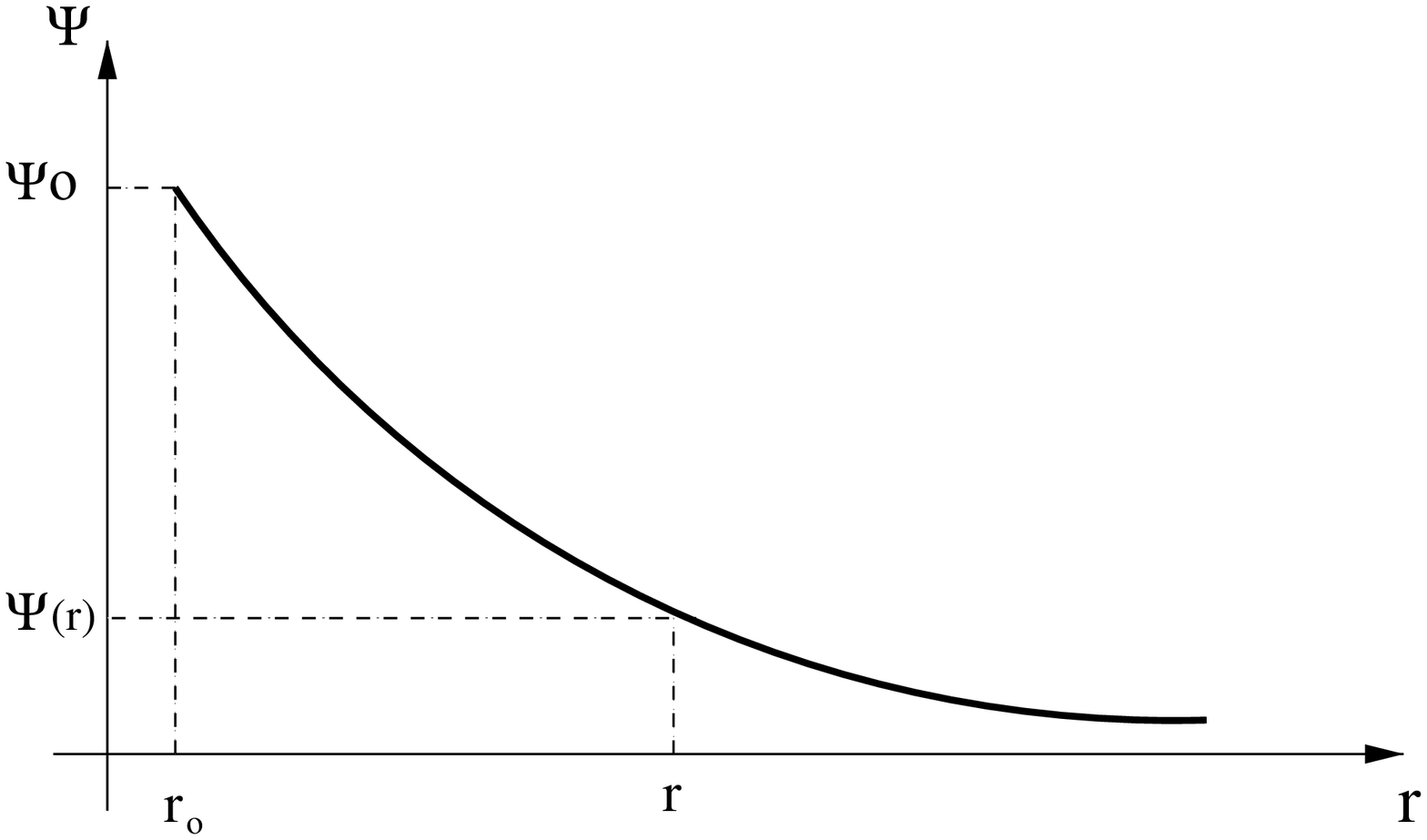} \caption{Case of an always decreasing proton potential energy. \label{emudecri}}
\end{figure}

\begin{figure} \plotone{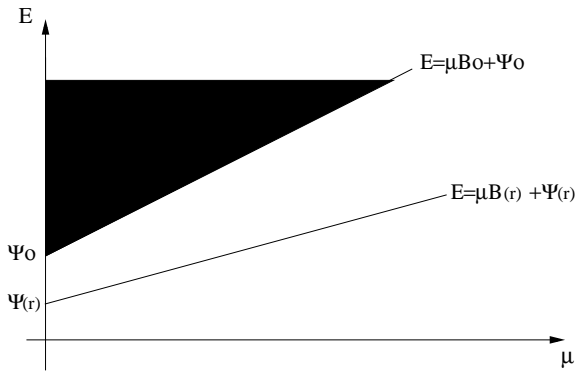} \caption{$E-\mu$ space for
the decreasing potential of Fig.\ref{emudecri}. All $r$ limit
lines are found below the $r_0$ one. The distribution function $f$
is defined in the black region for all altitudes $r\geq r_0$.
\label{emudecrii}}
\end{figure}

Let us now consider the case of a monotonic potential for the
protons, i.e. a monotonically decreasing potential, which is shown
in Figure $\ref{emudecri}$. All limit lines (shown in
Fig.$\ref{emudecrii}$) for all altitudes $r>r_0$ are below the
$r_0$ limit line and they never intersect each other (except for
$\mu<0$ which is an unphysical case). This is due to the fact that
both the limit line intercept and its slope are always decreasing.
At a given distance $r$, the distribution function $f$ is defined
above the corresponding $r$ limit line. But if the particles which
are present at an altitude $r$ are coming from the exobase $r_0$,
their function has also to be defined at the exobase, i.e. above
the $r_0$ limit line. In other words, the only region where the
VDF is defined in the $E-\mu$ space for particles coming from (or
getting back to) the exobase is above the $r_0$ limit line for all
the altitudes. The fact that the escaping particle region is
defined by only one limit line makes the case of a monotonic
potential particularly simple.

\begin{figure} \plotone{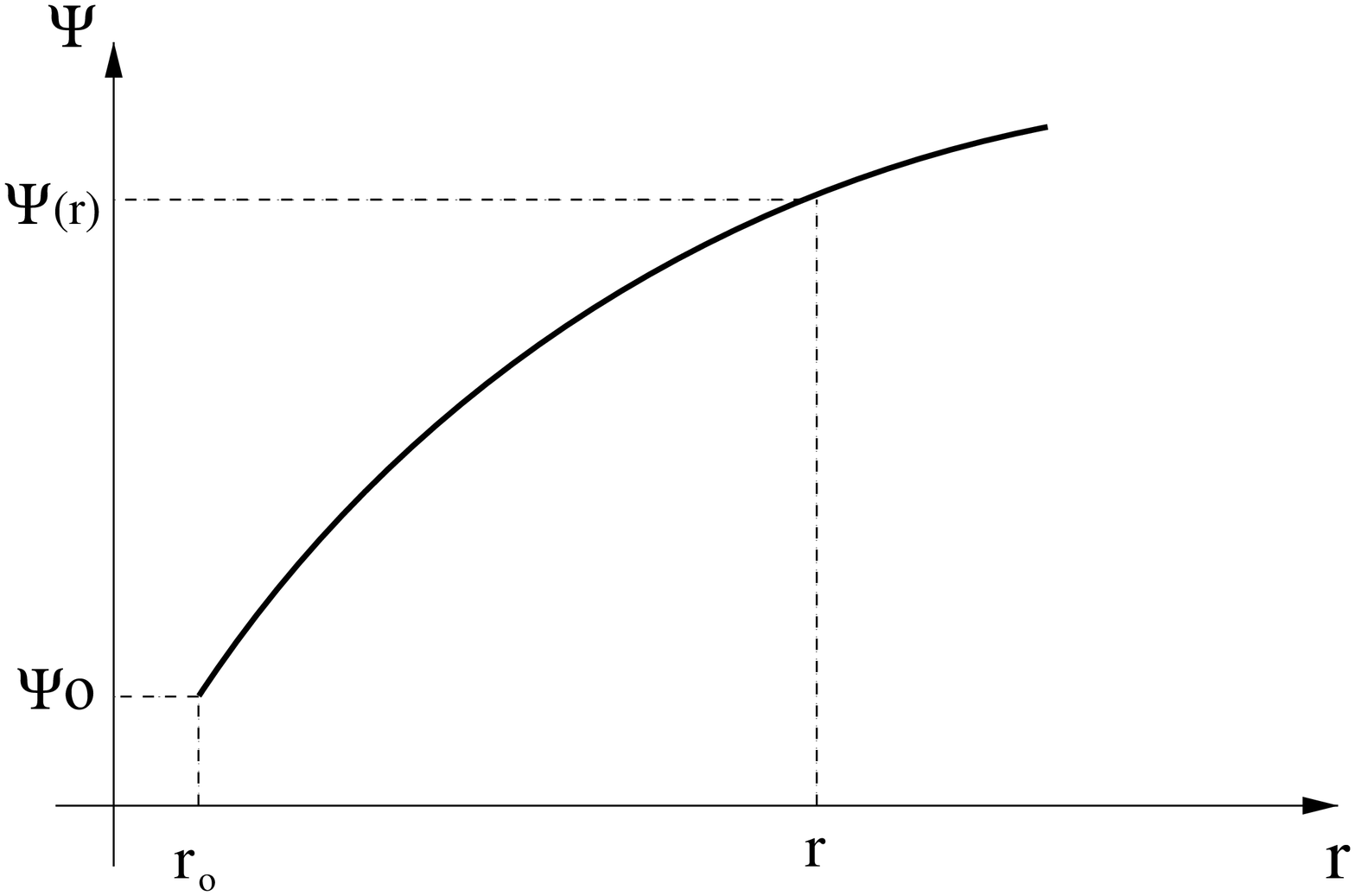} \caption{Case of an always increasing proton potential energy. \label{emuincri}}
\end{figure}

\begin{figure} \plotone{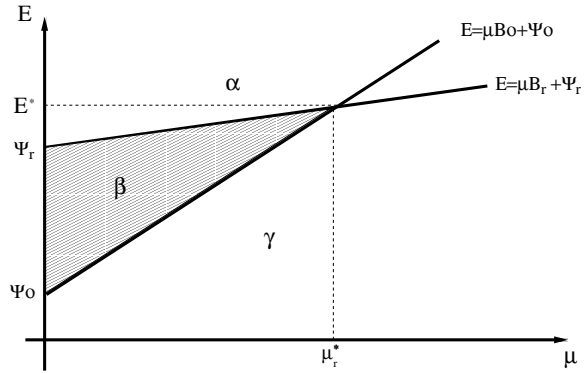} \caption{$E-\mu$ space for
the increasing potential of Fig.\ref{emuincri}. The $r$ limit
lines intersect. Protons that can escape from the altitude $r$ are
defined in the $\alpha$ region. The $\beta$ region corresponds to
ballistic protons, while the distribution function $f$ is not
defined in the $\gamma$ region. \label{emuincrii}}
\end{figure}

The opposite case of a monotonically increasing potential
(Fig.$\ref{emuincri}$) is slightly more complicated. The limit
lines slope is still decreasing, but the intercept is increasing,
giving the configuration of Figure $\ref{emuincrii}$. Now the
limit lines for the different altitudes intersect each other. If
we consider two altitudes $r_0$ and $r$, with $r>r_0$ (note that
$r$ could be the maximum altitude $r_{max}$ in the case of a
non-monotonic potential), the VDF $f$ is defined, as before, only
above the $r_0$ limit line and above the $r$ limit line for the
altitude $r$. This means that the escaping particles region is not
any more defined by only one line, but by two ones that intersect
at the point $(E^*,\mu^*_r)$ with
$\mu^*_r=-(\Psi_r-\Psi_0)/(B_r-B_0)$. This is the $\alpha$ region
shown in Figure $\ref{emuincrii}$. In the $\beta$ region, there
are particles which are defined in $r_0$, but are not present in
$r$, that is they are ballistic and can reach altitudes up to $r$
and then fall back to the exobase $r_0$.

\begin{figure} \plotone{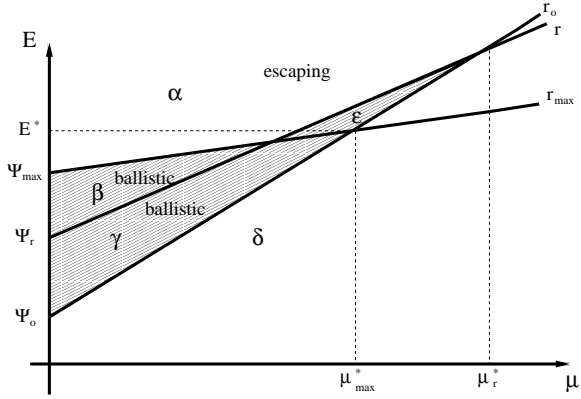} \caption{$E-\mu$ space for an increasing proton potential energy for altitudes
$r_0<r<r_{max}$, with $\mu^*_r>\mu^*_{max}$. The escaping protons
region $\alpha$ is reduced by the $\epsilon$ one, which contains
ballistic protons. \label{emubalistic}}
\end{figure}

\begin{figure} \plotone{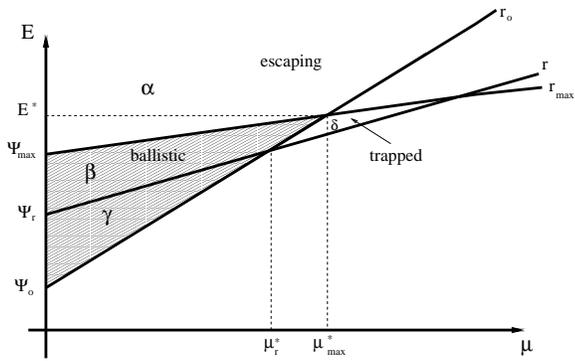} \caption{$E-\mu$ space for an increasing proton potential energy for altitudes
$r_0<r<r_{max}$, with $\mu^*_r<\mu^*_{max}$. The region $\delta$
corresponds to trapped particles between $r$ and $r_{max}$.
\label{emupieges}}
\end{figure}

In Figures $\ref{emubalistic}$ and $\ref{emupieges}$, we can see
the same case as before, considering the potential maximum at
$r_{max}$ and an intermediate altitude $r$, with $r_0<r<r_{max}$.
If $\mu^*_r$ is the abscissa of the intersection point between the
$r_0$ and $r$ limit lines, we can distinguish two different cases,
the first one when $\mu^*_r>\mu^*_{max}$, i.e. the $r$ limit line
is located above the intersection point $(E^*,\mu^*_{max})$
(Fig.$\ref{emubalistic}$) and the opposite one when
$\mu^*_r<\mu^*_{max}$ (Fig.$\ref{emupieges}$). In the first case,
the escaping region is not defined by only two lines as in Figure
$\ref{emuincrii}$, because these particles have also to be present
at altitude $r$. The consequence is that a new region appears, the
region $\epsilon$, which contains ballistic particles (between
$r_0$ and $r$) and not escaping ones. The totality of ballistic
particles between $r_0$ and $r$ are defined in region
$\gamma+\epsilon$. The $\beta$ region contains ballistic particles
that can reach altitudes up to $r_{max}$. We can therefore see
that the escaping particles region $\alpha$ is reduced (by the
$\epsilon$ region) due to the obligatory presence of these
particles in the intermediate altitude $r$ between the exobase and
$r_{max}$.

\begin{figure} \plotone{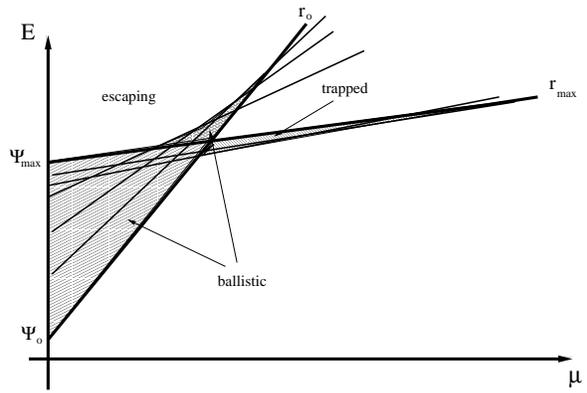} \caption{$E-\mu$ space for an increasing proton potential energy for altitudes
$r_0<r<r_{max}$ considering the contribution of several altitudes
between $r_0$ and $r_{max}$. \label{emutous}}
\end{figure}

In the second case of an increasing potential
(Fig.$\ref{emupieges}$), the escaping region is not modified, but
there is now a new region $(\delta)$ with trapped particles
between $r$ and $r_{max}$, which are defined below the $r_0$ limit
line and therefore do not come from the exobase. It is now evident
that definitions of the different species regions depend on the
potential values at all intermediate altitudes between $r_0$ and
$r_{max}$, as is shown in Figure ($\ref{emutous}$). The
geometrical definition of these regions is the main numerical
difficulty of our problem combined with the fact that the
potential values are not known but have to be calculated a
posteriori using the zero charge and current conditions.

\begin{figure} \plotone{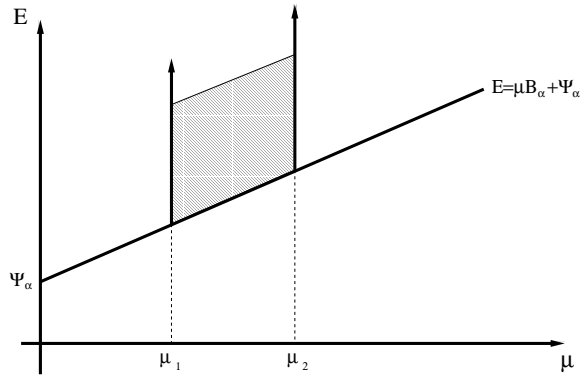} \caption{An example of an elementary region of escaping particles between two consecutive points $\mu_1$ and $\mu_2$ and above the $r_{\alpha}$ limit line.
\label{elementary}}
\end{figure}

At a given altitude $r_{\alpha}$, we can consider an elementary
region of escaping particles between two consecutive points
$\mu_1$ and $\mu_2$ and above the $r_{\alpha}$ limit line
(Fig.$\ref{elementary}$). In order to calculate a moment of the
VDF in a given altitude $r$ for the escaping particles, we have to
calculate the following integral:
\begin{eqnarray}\label{generintegral}
I_j(B_{\alpha},\Psi_{\alpha},\mu_1,\mu_2)=\int_{\mu_1}^{\mu_2}\left[\int_{E_{\alpha}}^{\infty}\hat{F}_j(E,\mu)dE\right]d\mu\nonumber
\end{eqnarray}
with $E_{\alpha}=\mu B_{\alpha}+\Psi_{\alpha}$. $\hat{F}_j$ is the
appropriate kernel from definitions
$(\ref{densemu})-(\ref{pressperpemu})$ with $B=B_r$ and
$\Psi=\Psi_r$. For a bi-Maxwellian VDF, these expressions have
been calculated by \citet{kha98} and will not be repeated here.
All different integrals $I_j$ have then to be added for all
altitudes $r_{\alpha}$ with $\alpha=0,...,m$. This has to be done
numerically.

Up to now, we have not found a general method for calculating the
exact solution of the self-consistent electric potential.
Therefore, we have to make an approximation on the escaping
particles rate in order to find a self-consistent potential. All
particles, which can be present at both $r_0$ and $r_{max}$, are
considered as escaping. i.e. we consider that the region labelled
$\epsilon$ in Figure $\ref{emubalistic}$ corresponds to escaping
particles at $r_m$, instead of ballistic which do not overcome
$r$. This important approximation is used by Jockers with a
resulting error in the particle flux claimed to be in general less
than 1$\%$ \citep{joc70}. However, the errors due to this
approximation may be much greater with suprathermal electrons.
With this approximation there are only three unknown parameters,
the position of the potential maximum $r_{max}$, the electric
potential $\phi_{E0}$ at the exobase and at the maximum
$\phi_{Emax}$. The main problem is that there are only two
equations (equality of fluxes at the exobase and quasi-neutrality
at $r_{max}$), but the problem has a unique solution if we suppose
the existence of only one maximum, which gives the following
constraints: $\phi_{Emax}<\phi_{E0}$,
$\Psi_{max}>\Psi_0$$\Rightarrow\phi_{Emax}>\phi_{E0}-(m_pM_\odot
G/e)(1/r_0-1/r_{max})$ and
$\Psi_{max}>0\Rightarrow$$\phi_{Emax}>m_pM_\odot G/(er_{max})$,
where $m_p$ is the proton mass. The technical details about the
solution of this system can be found in \citet{joc70} and equally
in \citet{lam03}. An important point to note is that the approach
used by \citet{lam03}, which consists in integrating the VDF in
the velocity space, is leading to exactly the same results as our
model, using the approximation on the $\epsilon$ region.

\begin{figure} \plotone{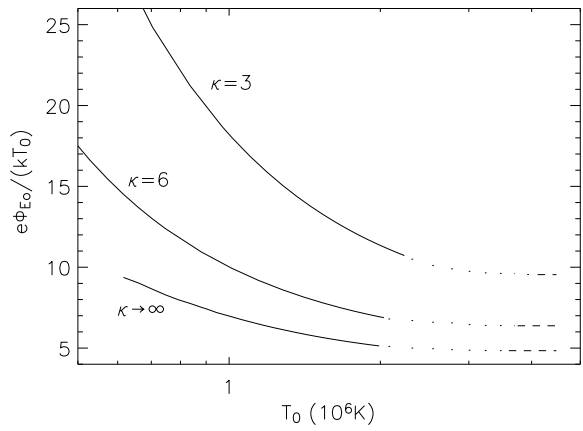} \caption{Electrostatic potential in function of the exobase temperature for different values of $\kappa$. Full lines represent solutions given by the present model (non-monotonic proton potential). Dashed lines correspond to previous exospheric models with a monotonic proton potential profile. Dotted lines show the range of a low numerical precision as explained in the text.\label{pottest}}
\end{figure}

In order to verify that the solutions given by the present model
are consistent with those of the previous models for which the
proton potential was monotonic \citep{mak97a}, we have calculated
the electrostatic potential in a range of exobase temperatures
including both types of solutions. For simplicity we assume for
this comparison a single Kappa VDF with $T_{p0}=T_{e0}=T_0$ and
take $r_0=5R_{\odot}$ in order to find solutions having a
monotonic potential in a reasonable temperature range. In Figure
\ref{pottest}, we can see the electrostatic potential $\phi_{E0}$
at the exobase (normalized to $k_bT_0/e$) as a function of $T_0$
for different values of $\kappa$. The full lines correspond to the
solutions given by the present model when the proton potential
profile is not monotonic. The dashed lines correspond to the
solutions obtained with a monotonic proton potential profile
\citep{mak97a}. These values are independent of the temperature
and represent the lowest possible acceleration obtained by
exospheric models. The dotted lines belong to the non-monotonic
regime as well, but the solutions are not known because of a low
precision due to the very small difference between $\phi_{E0}$ and
$\phi_{Emax}$. One can see that the solutions obtained in both
regimes are mutually consistent as there are no discontinuities.
The approximation used on the escaping particles rate would have
consequences only to the very left part of the curves in the
Figure \ref{pottest} for which the acceleration is important and
therefore the $\epsilon$ region would become significant. For the
same reasons the approximation made has more important
consequences for low values of $\kappa$. In a full treatment of
the problem that would not make this approximation, the calculated
terminal bulk speed would be higher. This is because protons in
the $\epsilon$ region would be considered as ballistic ones
(instead of escaping) and would thus need a stronger electrostatic
potential in order to escape from the gravitational well of the
Sun.

\clearpage


\clearpage
\clearpage

\end{document}